\documentclass[journal]{IEEEtran}
\usepackage{amsmath,amsfonts}
\usepackage{amsthm,amsmath,amssymb}
\usepackage{mathrsfs}
\usepackage{algorithmic}
\usepackage{algorithm}
\usepackage{array}
\usepackage{subfigure}
\usepackage{textcomp}
\usepackage{stfloats}
\usepackage{url}
\usepackage{verbatim}

\usepackage{graphicx}

\usepackage{cite}

\usepackage{bm}
\usepackage[shortlabels]{enumitem}
\usepackage{xcolor}
\usepackage{multirow}
\newtheorem{example}{\emph{\textbf{Example}}}
\renewcommand{\top}{T}

\begin{document}

\title{
 Partially Constrained GRAND of Linear Block Codes

\thanks{
 This work was supported by the National Key R\&D Program of China~(No. 2020YFB1807100). \emph{(Corresponding author: Xiao Ma.)}

 Yixin Wang is with the School of Systems Science and Engineering and the Guangdong Key Laboratory of Information Security Technology, Sun Yat-sen University, Guangzhou 510006, China (e-mail: wangyx58@mail2.sysu.edu.cn).
 
 Jifan Liang and Xiao Ma are with the School of Computer Science and Engineering and the Guangdong Key Laboratory of Information Security Technology, Sun Yat-sen University, Guangzhou 510006, China (e-mail: liangjf56@mail2.sysu.edu.cn; maxiao@mail.sysu.edu.cn).}
}

\author{%
 \IEEEauthorblockN{Yixin Wang, Jifan Liang, and Xiao Ma,~\IEEEmembership{Member,~IEEE}}
}



\maketitle

\begin{abstract}
\par This paper is concerned with a search-number-reduced guessing random additive noise decoding (GRAND) algorithm for linear block codes, called partially constrained GRAND (PC-GRAND). In contrast to the original GRAND, which guesses error patterns without constraints, the PC-GRAND guesses only those error patterns satisfying partial constraints of the codes. In particular, the PC-GRAND takes partial rows of the parity-check matrix as constraints for generating candidate error patterns and the remaining rows as checks for validating the candidates. The number of searches can be reduced when the serial list Viterbi algorithm (SLVA) is implemented for searching over a trellis specified by the partial parity-check matrix. This is confirmed by numerical results. Numerical simulations are also provided for comparison with other decoding algorithms.
\end{abstract}


\section{Introduction}
\par \IEEEPARstart{T}{he} maximum likelihood~(ML) decoding is optimal in terms of minimizing frame-error rate~(FER) when the prior distribution of the codewords is unknown or uniform. However, the high complexity of ML decoding makes it impractical for decoding a general code~\cite{Berlekamp1978NP}. Hence, the researchers mainly focus on practical near-ML decoders.
\par The ordered statistic decoding~(OSD) algorithm is a near-ML decoding algorithm~\cite{Fossorier1995OSD}. For a binary linear block code of dimension $k$ and minimum Hamming distance $d_\textrm{min}$, OSD can approach ML if order-$t$ reprocessing is implemented with $t=\lceil d_{\textrm{min}}/4-1\rceil$ but has a time complexity of $O(k^t)$. So the OSD is more promising for short block codes, and many efforts have been paid to reduce the complexity~\cite{ Wu2006Soft, Kabat2007OSD, Baldi2014OSD, Yue2019Segmentation,yue2022LEOSD,kim2022simplified}. Recently, a new variant of the OSD algorithm called locally constrained OSD~(LC-OSD) with much lower time complexity is investigated in~\cite{Wang2022LCOSD, Liang2023LCOSD}. Instead of order-$t$ reprocessing on the most reliable independent bits, the LC-OSD searches for test error patterns using the serial list Viterbi algorithm~(SLVA)~\cite{Nambirajan1994SLVA} over a trellis specified by a locally constrained parity-check matrix.
\par Motivated by the success of LC-OSD, we present a search-number-reduced guessing random additive noise decoding~(GRAND), called partially constrained GRAND~(PC-GRAND). The GRAND algorithm searches for the error patterns from the most likely to the least likely~\cite{Duffy2019GRAND}, which is ML. This idea was also mentioned in the introductory paragraph of~\cite{Valembois2001FPT}. The order of generating error patterns for GRAND can be specified in soft-GRAND~(SGRAND)~\cite{Solomon2020SGRAND} and ordered reliability bits GRAND~(ORBGRAND)~\cite{Duffy2022ORBGRAND}. Compared with the OSD algorithm, the GRAND algorithm does not need any matrix manipulations. If the decoding is successful, the resulting codeword for the GRAND algorithm is definitely an ML codeword, which cannot be guaranteed by the OSD algorithm. 
\par{The complexity of GRAND can be roughly measured by that of generating and checking a candidate error pattern multiplied by the number of searches. The conventional GRAND can generate one candidate in a simple way but has many unnecessary searches. One way to reduce the complexity is to reduce the number of searches at the expense of generating the candidates by imposing some constraints on the error patterns. In particular, several rows with disjoint non-zero positions are selected in [13] to limit the search space for ORBGRAND.  In this paper, we present a more general algorithm, called PC-GRAND, which guesses only those error patterns satisfying partial constraints of the linear block codes. Precisely, we partition the parity check matrix into two sub-matrices. One sub-matrix is used to generate candidate error patterns, while the other sub-matrix is used to check whether the searched pattern is valid. Distinguished from~\cite{Rowshan2022Constrianed}, the choice of the constraints is arbitrary, and the search is implemented by SLVA~\cite{Nambirajan1994SLVA} over the associated partially constrained trellis. Such an implementation over a trellis has at least two advantages. First, it provides a convenient way~(for those engineers familiar with trellis codes) to trade off the complexity, the throughput and the performance. Second, it provides a direct way to generalize the GRAND to the memory systems such as Markov noise channels~\cite{An2022GRANDMO}, intersymbol interference~(ISI) channels~\cite{Kavcic2003Interference} and trellis coded modulations.} Numerical results show that the PC-GRAND has less number of searches on average and can achieve the same performance as the LC-OSD algorithm for high-rate linear block codes.
\section{PC-GRAND}
\subsection{System Model}
Let $\mathbb{F}_2\triangleq\{0,1\}$ be the binary field and $\mathscr{C}[n, k]$ be a binary linear block code of length $n$ and dimension $k$.
Let $\mathbf{G}$ of size $k\times n$ be a generator matrix of $\mathscr{C}$ and $\mathbf{H}$ of size $(n-k)\times n$ be a parity-check matrix of $\mathscr{C}$. We have $\mathbf{G}\mathbf{H}^\top = \mathbf{O}$, where $\mathbf{O}$ is the all-zero matrix.
\par Let $\bm{u}=\{u_0,u_1,\cdots,u_{k-1}\}\in \mathbb{F}_2^k$ be an information vector to be transmitted. The information vector is first encoded into $\bm{c}=\bm u\mathbf{G}\in \mathbb{F}_2^n$ and then modulated by the binary phase shift keying~(BPSK) into a bipolar signal vector $\bm{x} \in \mathbb{R}^n$ as $x_i = 1-2c_i,~0\leq i< n$. Then the signal vector $\bm{x}$ is transmitted over an additive white Gaussian noise~(AWGN) channel, resulting in a received vector $\bm{y}\in\mathbb{R}^n$ given by $\bm{y} = \bm{x} + \bm{n}$, where $\bm{n}\sim \mathcal{N}(\bm{0}, \sigma^2\mathbf{I}_n)$ is a sample vector of white Gaussian noise. At the receiver, the hard-decision vector $\bm{z}\in\mathbb{F}_2^n$ is first delivered from the received vector $\bm y$ with

\begin{equation}
	z_i=
	\begin{cases} 
		0, & \mbox{if }y_i\geq0\\
		1, & \mbox{if }y_i<0\\
	\end{cases}
	,~0\leq i < n.
\end{equation}
The log-likelihood ratio~(LLR) vector, denoted by $\bm{r}\in\mathbb{R}^n$, is defined as
\begin{equation}
	\label{equ:calc-r}
	r_i = \log{\frac{p\{y_i\mid c_i=0\}}{p\{y_i\mid c_i=1\}}} = \frac{2y_i}{\sigma^2},~0\leq i < n,
\end{equation}
where $p\{\cdot\}$ is the (conditional) probability density function~(PDF).
We also refer~$|r_i|$ to as the reliability of $z_i$.
Also, the hypothetical error pattern $\bm{e}\in\mathbb{F}_2^n$ for a test codeword $\bm{v}\in\mathbb{F}_2^n$ is given by
	$
	\bm{e} = \bm{z} - \bm{v}.
	$
The ML decoding consists of finding a codeword $\bm{v}^*$ such that
\begin{equation}
	\bm{v}^* = \underset{\bm v\in \mathscr{C}}{\arg\max}~p\{\bm{y}\mid\bm{c}=\bm{v}\},
\end{equation}
which is equivalent to
\begin{equation}
	\bm{v}^* = \underset{\bm v\in \mathscr{C}}{\arg\min}~\log\frac{p\{\bm{y}\mid\bm{c}=\bm{z}\}}{p\{\bm{y}\mid\bm{c}=\bm{v}\}}.
\end{equation}
If we define the soft-weight of a test error pattern $\bm{e}$ by 
\begin{equation}
	\label{equ:soft-weight}
	\begin{split}
		\Gamma(\bm{e}) 
		= \log\frac{p\{\bm{y}\mid\bm{c}=\bm{z}\}}{p\{\bm{y}\mid\bm{c}=\bm{v}\}}
		= \log\frac{p\{\bm{y}\mid\bm{c}=\bm{z}\}}{p\{\bm{y}\mid\bm{c}=\bm{z}-\bm{e}\}} \\
		= \sum_{i=0}^{n-1} \log{\frac{p\{y_i\mid c_i=z_i\}}{p\{y_i\mid c_i=z_i-e_i\}}}
		= \sum_{i=0}^{n-1} e_i|r_i|,
	\end{split}
\end{equation}
we see that the ML decoding is equivalent to the lightest-soft-weight~(LSW) decoding.
\subsection{SGRAND}
\par The SGRAND is a soft detection ML decoder~\cite{Solomon2020SGRAND}. The decoder first sorts the bits in the hard-decision vector $\bm z$ in ascending order according to their reliabilities, resulting in~$\widetilde{\bm z}$. The error patterns are searched according to the reliabilities of bits with a maximum searching number $\ell_{\textrm {max}}$. At each search step, the pattern ${\bm e}$ with the lightest soft weight will be chosen and removed from a priority queue $S$. The searched pattern ${\bm e}$ is then used to check whether $\bm z-{\bm e}$ is a valid codeword with the parity check matrix $\mathbf{H}$. In the case when $\bm e$ does not satisfy the checks, the successors of ${\bm e}$ will be inserted into $S$. The details of the SGRAND can be found in~\cite{Solomon2020SGRAND}. Notice that the procedure specified in~\cite[Algorithm~2]{Solomon2020SGRAND} to generate the ordered error patterns can be described with a flipping pattern tree~(FPT)~\cite{lin2020thesis}\footnote{The description with FPT in~\cite{lin2020thesis} is more general and applicable to nonbinary codes.}.
\subsection{PC-GRAND}
In this subsection, we present the PC-GRAND algorithm. 
\subsubsection{Preprocessing} Let $\delta$ be an integer such that $0\leq \delta\leq n-k$. Divide the parity check matrix $\mathbf{H}$ into two sub-matrices denoted as 
\begin{equation}
	\label{decompose1}
	\mathbf{H}=\begin{bmatrix}
		\mathbf{H}_1\\
		\mathbf{H}_2\\
	\end{bmatrix},
\end{equation}
where the sub-matrix $\mathbf{H}_1$ is of size $\delta\times n$ and the sub-matrix $\mathbf{H}_2$ is of size $(n-k-\delta)\times n$. 
If a test vector $\bm v$ is a valid codeword, we have 

\begin{equation}
	{\mathbf{H}}\bm v^\top={\mathbf{H}}(\bm z^\top-\bm e^\top)=\bm 0.
\end{equation} 
Equivalently, we have both
 \begin{equation}
	\label{equ:search_e-parity}
	{\mathbf{H}}_1\bm e^\top={\mathbf{H}}_1\bm z^\top
\end{equation}
and
\begin{equation}
		\label{equ:check_e-parity}
	\mathbf{H}_2\bm e^\top=\mathbf{H}_2\bm z^\top	.
\end{equation}
Upon receiving $\bm y$, the hard-decision vector $\bm z$ is determined, and the right-hand side~(RHS) of \eqref{equ:search_e-parity} and~\eqref{equ:check_e-parity} are calculated and stored for future use.
\subsubsection{Search Scheme}
We can use ~\eqref{equ:search_e-parity} as constraints to search for candidate error patterns with the soft weights in non-decreasing order, i.e., $\Gamma(\bm e^{(i)})\leq \Gamma(\bm e^{(i+1)})$, which can be achieved by the SLVA~\cite{Nambirajan1994SLVA} over a trellis specified by the sub-matrix ${\mathbf{H}}_1$. For every searched candidate pattern $\bm e^{(i)}$, the equation~\eqref{equ:check_e-parity} is used to check whether the pattern $\bm e^{(i)}$ is valid. The search scheme is described in Algorithm~\ref{algo:coinstrian_GRAND} by setting the maximum number of searches, $\ell_{ \textrm{max}}$.


\begin{algorithm}[t]
	\renewcommand{\algorithmicrequire}{\textbf{Input:}}
	\renewcommand{\algorithmicensure}{\textbf{Output:}}
	\caption{Search Scheme for PC-GRAND}
	\label{algo:coinstrian_GRAND}
	\begin{algorithmic}[1] 
		\REQUIRE $\bm{z}$, $\delta$, $\ell_{\textrm {max}}$.
		\ENSURE the optimal searched codeword $\bm{v}$.
		\STATE Perform preprocessing.
		\STATE ${{\bm{e}}}_{\textrm{opt}} \leftarrow {\bm{0}}$
		\FOR{$i= 1, 2, 3, \ldots, \ell_{\textrm {max}}$}
		\STATE Find the $i$-th lightest-soft-weight test error pattern $\bm e^{(i)}$ such that ${\mathbf{H}}_1 (\bm e^{(i)})^{\top} = {\mathbf{H}}_1 \bm z^{\top}$.
		\label{line:LC-OSD:find-jth}
		\IF{$\mathbf{H}_2 (\bm e^{(i)})^{\top} = \mathbf{H}_2 \bm z^{\top}$}
		\STATE $\bm e_{\textrm {opt}}\leftarrow\bm e^{(i)}$
		\STATE \textbf{break}
		\ENDIF
		\ENDFOR
		\STATE $\bm{v} \leftarrow ({{\bm{z}} - {\bm e_{\textrm {opt}}}})$.
		\RETURN $\bm{v}$
	\end{algorithmic}
\end{algorithm}

\section{Complexity Analysis}

\par We analyze the computational complexity of the decoding algorithm by evaluating the number of floating point operations~(FLOPs) and binary operations~(BOPs) of each step. Denote by $\ell_{\textrm{avg}}$ the average number of test patterns per received vector. 
\subsection{Computational Complexity}
\par We first analyze the complexity for SGRAND~\cite{Solomon2020SGRAND}. The sorting requires $\mathcal{O}(n\log n)$ BOPs and FLOPs. {In the case when a candidate error pattern $\bm e$ does not pass the parity checks, two new error patterns are generally needed to be inserted into the queue $S$. One new pattern $\bm e_1$ can be immediately generated from $\bm e$ by flipping one bit with $\mathcal{O}(1)$ BOPs. The other new error pattern is then generated by copying $\bm e_1$ and flipping one bit with $\mathcal{O}(n)$ BOPs.} The store of the error patterns can be implemented with the min-heap, and hence the insertion and deletion complexity of new patterns can be upper bounded by $\ell_{\textrm{avg}}\log\ell_{\textrm{avg}}$ FLOPs. The checking is the multiplication of the matrix and the vector, which requires $\mathcal{O}(n(n-k))$ BOPs. Thus, the computational complexity of SGRAND can be evaluated as
\begin{equation}
	\begin{aligned}
		T_{\textrm{avg}}
		&= \underbrace{\mathcal{O}(n\log n)}_{\textrm{sorting}}
		+ \underbrace{\mathcal{O}(\ell_{\textrm{avg}}(n+\log\ell_{\textrm{avg}}))}_\textrm{searching} 
		+ \underbrace{\mathcal{O}({\ell}_\textrm{avg}n(n-k))}_\textrm{checking}.
	\end{aligned}
\end{equation}
\par For the proposed PC-GRAND, the trellis specified by the local parity-check matrix ${\mathbf{H}}_1$ has $n$ sections, and each section has at most $2^\delta$ states. To find the best path ${{\bm{e}}}^{(1)}$, the SLVA needs to calculate and store the best paths associated with all allowable states, each requiring $\mathcal{O}(1)$ BOPs and FLOPs. Thus, for finding the best path, the complexity is $\mathcal{O}(2^\delta\cdot n)$. With the previous $i-1$ paths found, searching for a candidate ${{\bm{e}}}^{(i)}\,(i >1)$ by the SLVA requires $\mathcal{O}(n)$ FLOPs. Upon generating a candidate error pattern, checking with $\mathbf{H}_2$ requires $\mathcal{O}(n(n-k))$ BOPs. Thus, the complexity of PC-GRAND can be evaluated as
\begin{equation}
	\begin{aligned}
		T_{\textrm{avg}}
		&= \underbrace{\mathcal{O}(2^\delta\cdot n+\ell_{\textrm{avg}}n)}_\textrm{searching over trellis} + \underbrace{\mathcal{O}({\ell}_\textrm{avg}n(n-k))}_\textrm{checking}.
	\end{aligned}
\end{equation}
\par {From the analysis above, we see that the time complexity is dominated by the average search number $\ell_{\textrm{avg}}$. In the case when the $\ell_{\textrm{avg}}$ of PC-GRAND is far less than that of SGRAND, the time complexity for PC-GRAND can be lower than SGRAND. }
\subsection{Space Complexity}
 {We analyze the space complexity of the algorithms for the searching in PC-GRAND and SGRAND by calculating the number of bytes used.}

{In the searching for SGRAND, upon selecting a candidate error pattern ${\bm e}$ in $S$, at most two new patterns generated from ${\bm e}$ will be inserted into $S$. Thus, the main storage space for SGRAND is to store the set $S$ of size $\mathcal{O}(\ell_{\textrm{max}}n)$. The main storage space of SLVA includes the space occupied by the initialization, $\mathcal{O}(2^\delta\cdot n)$ and the space occupied by searching and storing paths, $\mathcal{O}(\ell_{\textrm{max}}\cdot n)$. In summary, the space complexity for PC-GRAND is given by $\mathcal{O}((2^\delta+\ell_{\textrm{max}})n)$. From the analysis above, if the value of $\delta$ and $2^{\delta}$ is far less than $\ell_{\rm max}$, the space complexity of PC-GRAND and SGRAND  is almost the same. Actually, to achieve the same performance, the $\ell_{\rm max}$ for PC-GRAND can be less than that for the SGRAND.
}

\section{Simulation Results}
In this section, we present simulation results to demonstrate the performance of the PC-GRAND algorithm. The SGRAND algorithm~\cite{Solomon2020SGRAND} and LC-OSD algorithm~\cite{Liang2023LCOSD} are also implemented for comparison. 
\par We denote by $\ell_{\textrm{avg}}$ and $\ell_{\textrm{max}}$, respectively, the average search number and the maximum search number for PC-GRAND, LC-OSD and SGRAND. We denote by $\delta$ the number of partial constraints for PC-GRAND and local constraints for LC-OSD~\cite{Liang2023LCOSD}.
\begin{figure}[t]
	\centering
	\subfigure[The FER ]{\includegraphics[width=0.4\textwidth]{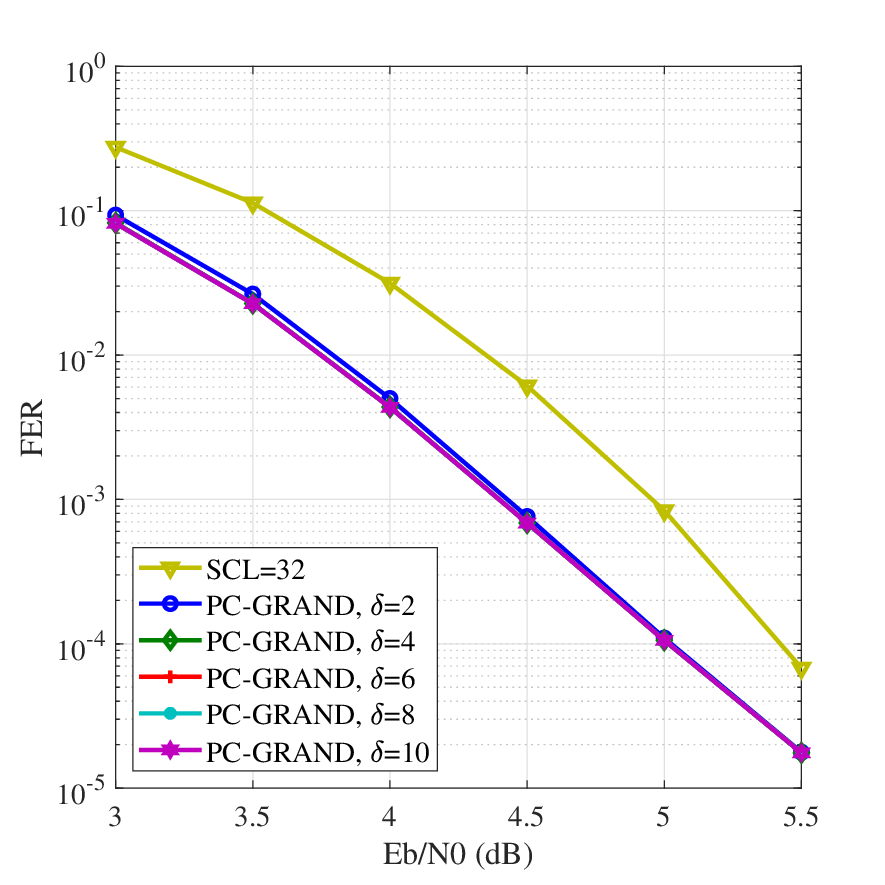}\label{fig:CA_polar_delta}}
	\hspace{-0.2cm}
	\subfigure[The average search number $\ell_{\textrm{avg}}$ ]{\includegraphics[width=0.4\textwidth]{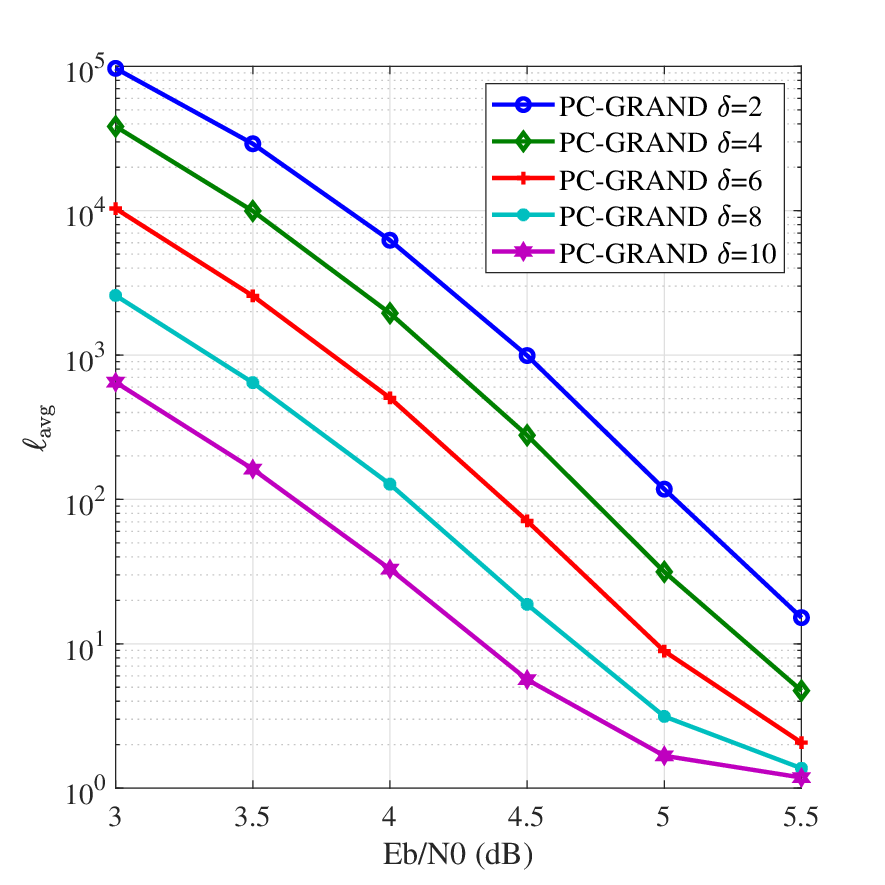}\label{fig:avg_search}}
	\caption{	The FER and the average search number $\ell_{\textrm{avg}}$for CA-polar $\mathscr{C}[128,105]$ with different $\delta$ in PC-GRAND. The maximum search number is $\ell_{\textrm{max}}=10^6$. }
\end{figure}


\begin{example}\rm
	In this example, we consider the cyclic redundancy check~(CRC)-aided polar~(CA-polar) code $\mathscr{C}[128,105]$ in 5G new ratio~(NR) for uplink communication. The simulation results for FER with different $\delta$ for PC-GRAND with $\ell_{\textrm{max}}=10^6$ are shown in Fig.~\ref{fig:CA_polar_delta}. The performance of the CRC-aided successive cancellation list decoding~(CA-SCL)~\cite{Niu2012CASCL, Tal2015CASCL} with list size~$32$ is also shown in this figure. From Fig.~\ref{fig:CA_polar_delta}, we can see that, for all values of $\delta$, {the PC-GRAND outperforms the SCL~(with list size~32), achieving a gain 0.3~dB at FER $\approx 10^{-4}$.} Generally, the FER can be improved as $\delta$ grows. Also notice that, the performance of FER can hardly be improved as $\delta$ grows when $\delta>6$.
	\par We also investigate the average search number $\ell_{\textrm{avg}}$ for different $\delta$ and the simulation results are shown in Fig.~\ref{fig:avg_search}. We can observe from Fig.~\ref{fig:avg_search} that $\ell_{\textrm{avg}}$ can be reduced as the partial constraints of the SLVA increase. To make a trade-off between performance and complexity, we choose $\delta=6$ for PC-GRAND in the following comparison.
\end{example}


\begin{figure}[t]
	\centering

	\subfigure[ CA-polar]{\includegraphics[width=0.4\textwidth]{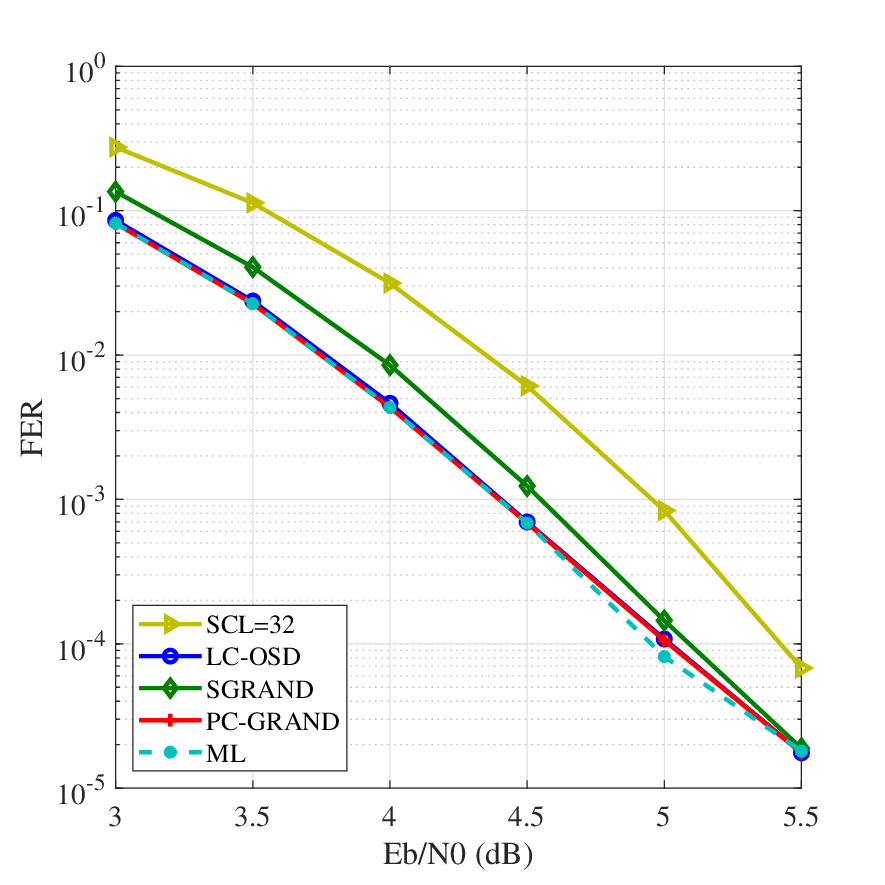}\label{fig:CA_polar_comparison}}
	\hspace{-0.2cm}
	\subfigure[BCH]{\includegraphics[width=0.4\textwidth]{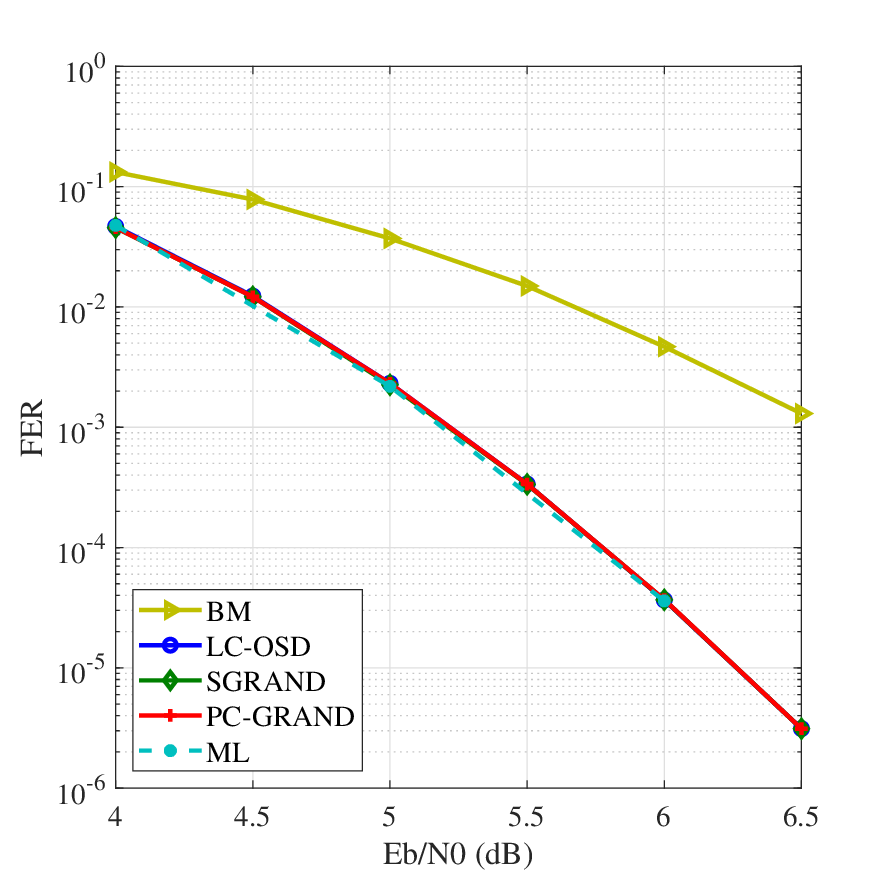}\label{fig:BCH_comparison}}
	\caption{	The FER of CA-polar code $\mathscr{C}[128,105]$ and BCH code $\mathscr{C}[127,113]$ for PC-GRAND ($\delta=6, \ell_{\textrm {max}}=10^6$), LC-OSD~\cite{Liang2023LCOSD} ($\delta=8, \ell_{\textrm {max}}=16384$) and SGRAND~\cite{Solomon2020SGRAND} ($\ell_{\textrm {max}}=10^6$). The ML simulation results~\cite{Liang2023LCOSD} are also plotted. }
		\label{fig:FER}
\end{figure}

%
\begin{example}\rm 
We compare the performance of PC-GRAND~($\delta=6, \ell_{\textrm {max}}=10^6$), LC-OSD~\cite{Liang2023LCOSD}~($\delta=8$, $\ell_{\textrm {max}}=16384$) and SGRAND~\cite{Solomon2020SGRAND}~($\ell_{\textrm {max}}=10^6$) for CA-polar code $\mathscr{C}[128,105]$ and Bose-Chaudhuri-Hocquenghem (BCH) code $\mathscr{C}[127,113]$. The simulation results are shown in Fig.~\ref{fig:FER}.  From Fig.~\ref{fig:FER}, we see that all the three algorithms can approach the ML performance. For CA-polar code, PC-GRAND performs slightly better than SGRAND. This is because SGRAND with $\ell_{\textrm{max}}$ is set to abandon before it identifies the ML codeword while PC-GRAND can identify the ML codeword in fewer queries with the constraints allowing it to skip invalid error patterns. 

\end{example}


\begin{figure}[t]
	\centering
	\subfigure[ CA-polar]{\includegraphics[width=0.4\textwidth]{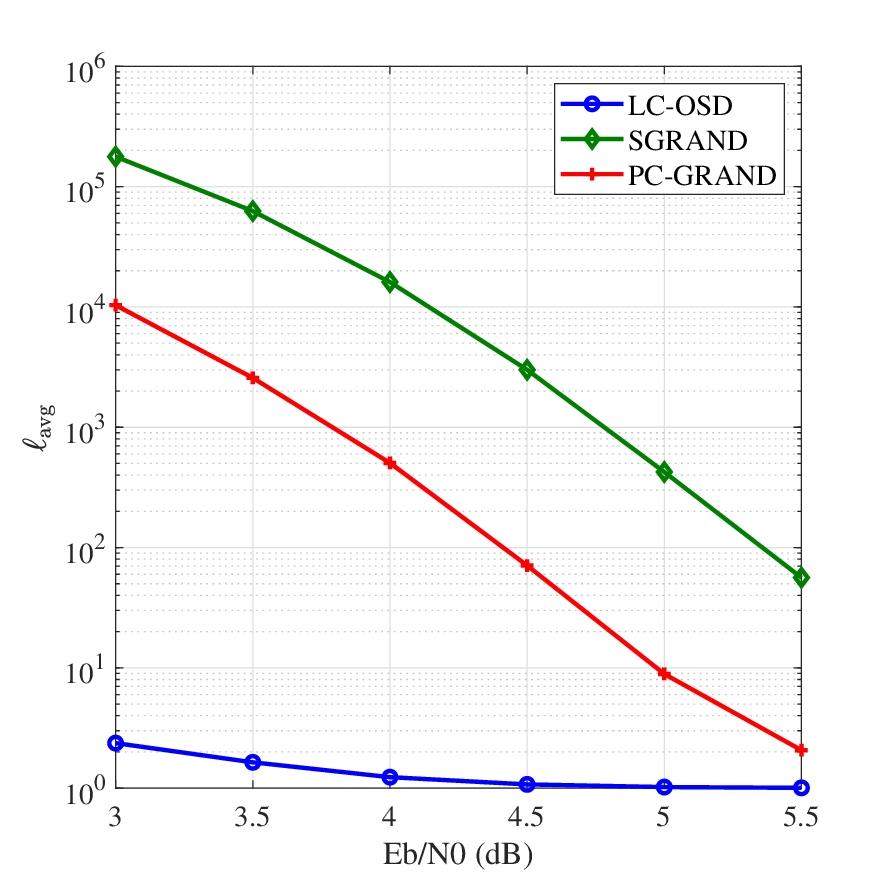}
		\label{fig:avg_search_comparison}}
	\hspace{-0.2cm}
	\subfigure[BCH]{\includegraphics[width=0.4\textwidth]{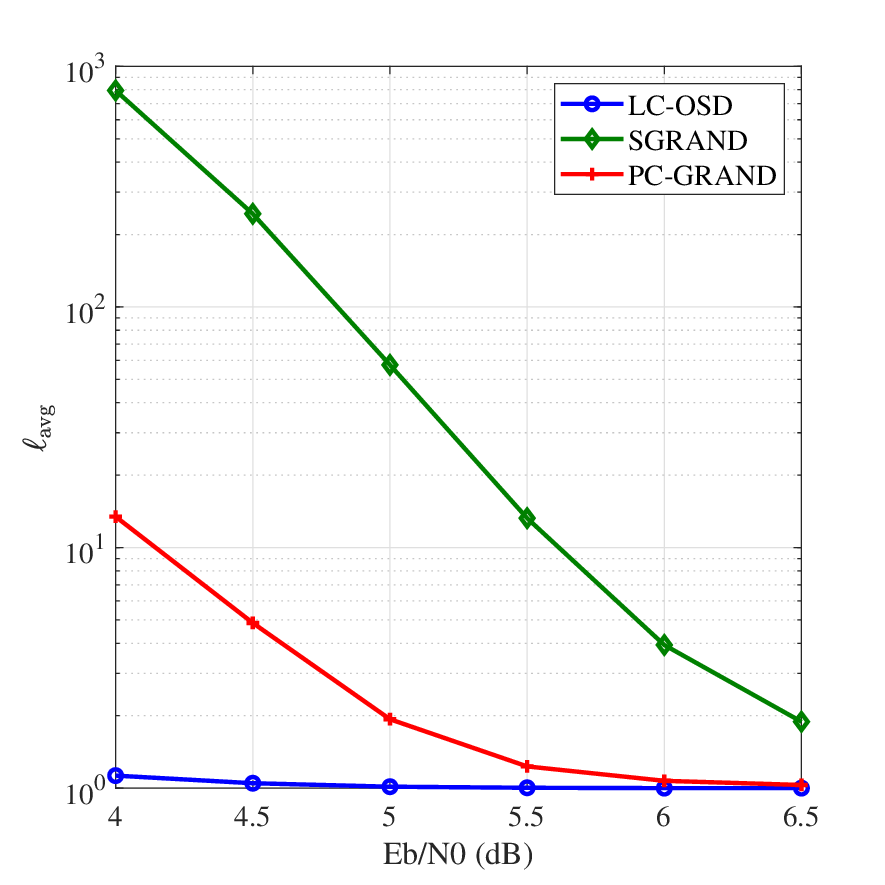}
		\label{fig:BCH_search_comparison}}
	\caption{The average search number of CA-polar code $\mathscr{C}[128,105]$ and BCH code $\mathscr{C}[127,113]$ for PC-GRAND ($\delta=6, \ell_{\textrm {max}}=10^6$), LC-OSD~\cite{Liang2023LCOSD} ($\delta=8, \ell_{\textrm {max}}=16384$) and SGRAND~\cite{Solomon2020SGRAND} ($\ell_{\textrm {max}}=10^6$).}
	\label{fig:search}
\end{figure}

\begin{figure}[t]
	\centering
	\subfigure[ CA-polar]{\includegraphics[width=0.4\textwidth]{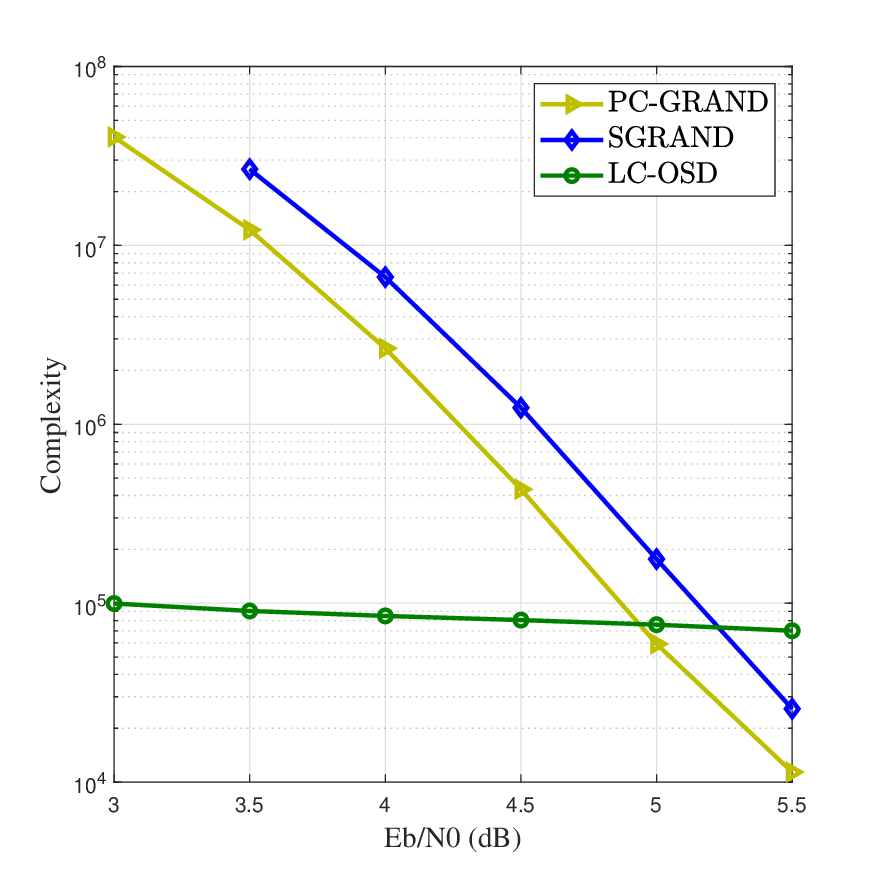}
		\label{fig:complexity_comparison_ca_polar}}
	\hspace{-0.2cm}
	\subfigure[BCH]{\includegraphics[width=0.4\textwidth]{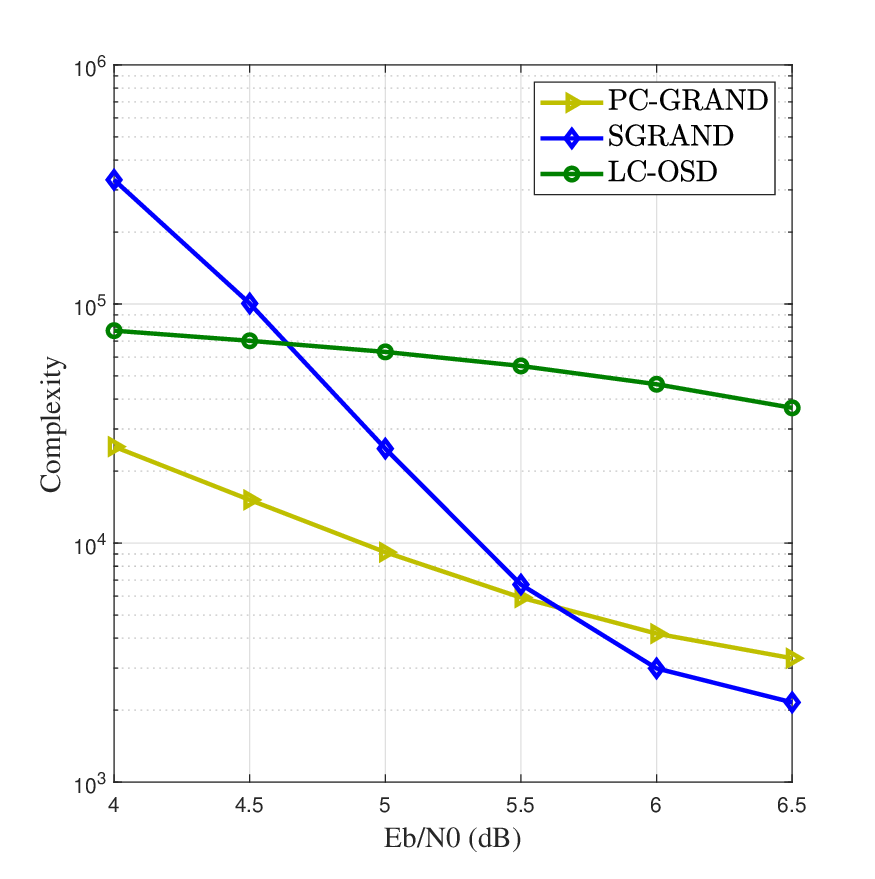}
		\label{fig:complexity_comparison_BCH}}
	\caption{The  BOPs and FLOPs in software for CA-polar code $\mathscr{C}[128,105]$ and BCH code $\mathscr{C}[127,113]$ for PC-GRAND ($\delta=6, \ell_{\textrm {max}}=10^6$), LC-OSD~\cite{Liang2023LCOSD} ($\delta=8, \ell_{\textrm {max}}=16384$) and SGRAND~\cite{Solomon2020SGRAND} ($\ell_{\textrm {max}}=10^6$).}
	\label{fig:complexity}
\end{figure}

\begin{example}\rm
	 {In this example, we show the computational and space complexity of the decoding algorithms. Since the computational complexity is dominated by the average search number $\ell_{\rm avg}$, we first compare the average search number of PC-GRAND~($\delta=6, \ell_{\textrm {max}}=10^6$), LC-OSD~\cite{Liang2023LCOSD}~($\delta=8$, $\ell_{\textrm {max}}=16384$) and SGRAND~\cite{Solomon2020SGRAND}~($\ell_{\textrm {max}}=10^6$) for CA-polar code $\mathscr{C}[128,105]$ and BCH code $\mathscr{C}[127,113]$. The simulation results are shown in  Fig.~\ref{fig:search}.  We also count the BOPs and FLOPs on average needed to decode a codeword in software implementation for the three algorithms and the results are shown in Fig.~\ref{fig:complexity}. With the results in Figs.~\ref{fig:search} and~\ref{fig:complexity}, we see that, the LC-OSD has the least $\ell_{\textrm{avg}}$ but requires Gaussian elimination for preprocessing with computational complexity of almost $\mathcal{O}(n^3)$ for decoding every noisy codeword. Thus, LC-OSD can have high decoding complexity in high SNR region, compared with SGRAND and PC-GRAND.  PC-GRAND can significantly reduce the average search number, compared with SGRAND, and hence can have less average computational complexity for decoding some codes.  For the simulations, we set $\ell_{\rm max}$  for both SGRAND and PC-GRAND to $10^6$. With this setting, $2^\delta$ is far less than $\ell_{\rm max}$  and the space complexity for SGRAND and PC-GRAND is almost the same.}
\end{example}
\section{Conclusion}
In this paper, we have proposed the PC-GRAND to reduce the average search number of SGRAND and hence can reduce the decoding complexity. More specifically, a small number of rows from the parity check matrix are used to constrain the candidate error pattern search in SLVA over an associated partially constrained trellis. The remaining rows are used as checks for validating the candidates. The computational complexity analysis and numerical results show that introducing partial constraints can reduce decoding complexity compared with SGRAND. The comparison results show that the PC-GRAND performs the same as LC-OSD. {Since the PC-GRAND is implemented over a trellis, it can be easily generalized to memory channels. { Although it is presented as a serial algorithm over a trellis in this paper, the PC-GRAND
		can be implemented in parallel by partitioning the trellis into multiple sub-trellises and performing separately the SLVA over each sub-trellis simultaneously.}}

\bibliographystyle{IEEEtran}
\bibliography{ref}

\end{document}